\documentclass[english,draft]{aipproc}
\usepackage[T1]{fontenc}
\usepackage[latin1]{inputenc}
\usepackage{graphicx}

\makeatletter

\layoutstyle{6x9}

\usepackage{babel}
\makeatother
\begin{document}

\newcommand{\missET}{{E_{T}\!\!\!\!\!\!\!/\:\:\,}}

\newcommand\gsim{\mathrel{\rlap{\raise.4ex\hbox{$>$}} {\lower.6ex\hbox{$\sim$}}}}   \newcommand\lsim{\mathrel{\rlap{\raise.4ex\hbox{$<$}} {\lower.6ex\hbox{$\sim$}}}}   

\def\gtrsim{\gsim} 

\def\lesssim{\lsim} 

\def\alphas{\alpha_s }

\def\D0{D\O~} 

\def\MSbar{$\overline{{\rm MS}}$}

\def\Msbar{\MSbar}

\def\msbar{\MSbar}



\classification{} 


\author{Stefan Kretzer}{ address={Physics Department, Brookhaven National Laboratory, Upton, New York 11973}, altaddress={ RIKEN-BNL Research Center,Brookhaven National Laboratory, Upton, New York 11973 -- 5000 } } 

\author{Fredrick I. Olness}{ address={Department of Physics, Southern Methodist University, Dallas, Texas 75275-0175} }

\begin{flushright}hep-ph/0508216 \end{flushright}

\title{Heavy Quark Parton Distribution Functions\thanks{The research
presented here was performed in collaboration with W.K.~Tung,
P.~Nadolsky, J.F.~Owens, J.~Pumplin, D.~Stump, J.~Huston, \&
H.L.~Lai.},\ \thanks{Contribution to the proceedings of the XIII
International Workshop on Deep Inelastic Scattering (DIS 2005), April
27 - May 1, 2005, Madison, WI, U.S.A.  Presented by Fredrick Olness.}}

\begin{abstract}
We present the CTEQ6HQ parton distribution set which is determined
in the general variable flavor number scheme which incorporates heavy
flavor mass effects; hence, this set provides advantages for precision
observables which are sensitive to charm and bottom quark masses.
We describe the analysis procedure, examine the predominant features
of the new distributions, and compare with previous distributions.
We also examine the uncertainties of the strange quark distribution
and how the the recent NuTeV dimuon data constrains this quantity.
\end{abstract}
\maketitle
\title{Heavy Quark Parton Distribution Functions}

Parton distributions functions (PDFs) provide the essential link between
the theoretically calculated partonic cross-sections, and the experimentally
measured physical cross-sections involving hadrons and mesons. This
link is crucial if we are to make incisive tests of the standard model,
and search for subtle deviations which might signal new physics. The
choice of the renormalization scheme is an important issue to address
if we are to make the most efficient use of our fixed-order perturbation
expansion. Separately, it is important to know the uncertainty range
of the PDFs, and properly fold these into the overall uncertainty
estimates. We will address both of these issues in turn.

\subsubsection{\textsc{Global Analysis and the CTEQ6HQ PDFs:} }

The CTEQ6HQ (or C6HQ for short) PDFs\cite{Kretzer:2003it} are obtained
by performing a global analysis using the generalized (non-zero quark-mass)
\Msbar\ perturbative QCD framework of Refs.~\cite{Aivazis:1993pi,Tung:2001mv},
which we label the general-mass variable-flavor-number scheme (GM-VFNS).
When matched to the corresponding hard-scattering cross-sections calculated
in the same scheme, the combination should provide a more accurate
description of the precision DIS structure function data, as well
as other processes which are sensitive to charm and bottom mass effects. 

The C6HQ global fitting follows the same procedure as that of the
earlier CTEQ6 analysis.\cite{Pumplin:2002vw} The data sets used before
are supplemented by the H1 and ZEUS data sets for the structure function
$F_{2}^{c}$ with tagged charm particles in the final state. The $F_{2}^{c}$
data sets are quite relevant for this analysis since $F_{2}^{c}$
is sensitive to the charm and gluon distributions, which are tightly
coupled in the generalized \Msbar\
formalism. 

The C6HQ set is the best fit obtained with these inputs. A broad measure
of the quality of this fit is provided by the overall $\chi^{2}$
of 2008 for a total number of 1925 data points ($\chi^{2}$/DOF =
1.04). This is to be compared to a $\chi^{2}$ of 2037 for 1925 points
($\chi^{2}$/DOF = 1.07) in the case of CTEQ6M (or C6M for short).
The new C6HQ fit reduces the overall $\chi^{2}$ by 29 out of $\sim$2000
as compared to the C6M fit. The improvement of this generalized \Msbar\ result
over the zero-mass \Msbar\ result is encouraging, since the generalized
\Msbar\ formalism represents a more accurate formulation of PQCD.
However, a difference of $\chi^{2}$ of 29 is within the current estimated
range of uncertainty of PDF analysis. \cite{Pumplin:2002vw} Therefore,
the significance of this difference is arguable. We also note that
the improvement in $\chi^{2}$ is spread over most of the data sets:
there is no {}``smoking gun'' for the overall difference.

Since perturbative calculations are renormalization scheme dependent,
it is important to use properly matched hard-scattering cross-sections
and PDFs when evaluating factorized cross-sections for physical applications.
This issue is particularly relevant for applications involving heavy
quarks, since the heavy quark introduces a new mass scale which leads
to complications of the PQCD formalism. To illustrate this point,
we compare the above results with two possible uses of the PDFs that
represent a \emph{mis-use} of PQCD in principle, but occur frequently
in the literature in practice, perhaps out of necessity. These involve
using PDFs obtained in the general-mass scheme convoluted with hard-scattering
cross-sections (Wilson coefficients) defined in the zero-mass scheme,
and vice versa. For example, if we use the C6M PDFs which are derived
in the ZM-VFN scheme with the GM-VFN hard-scattering cross-sections,
we obtain a total $\chi^{2}$ of 2431 for 1925 data points ($\chi^{2}$/DOF
= 1.26). Conversely, if we use the C6HQ PDFs which are derived in
the GM-VFN scheme with the ZM-VFN hard-scattering cross-sections,
we obtain a total $\chi^{2}$ of 2496 for 1925 data points ($\chi^{2}$/DOF
= 1.30). For the same data sets, these mis-matched schemes result
in a overall $\chi^{2}$ difference of 420$\sim$490 compared to the
C6HQ set ($\chi^{2}$ of 2008). These are quite large differences
relative to the tolerances discussed in Refs.~\cite{Martin:2000wq,Martin:2002aw,Martin:2003sk,Pumplin:2002vw},
and result in obvious discrepancies with some of the precision DIS
data sets. Clearly, \emph{}for quantitative applications it is imperative
to maintain consistency between the PDFs and the hard-scattering cross-sections\emph{.}

\begin{figure}
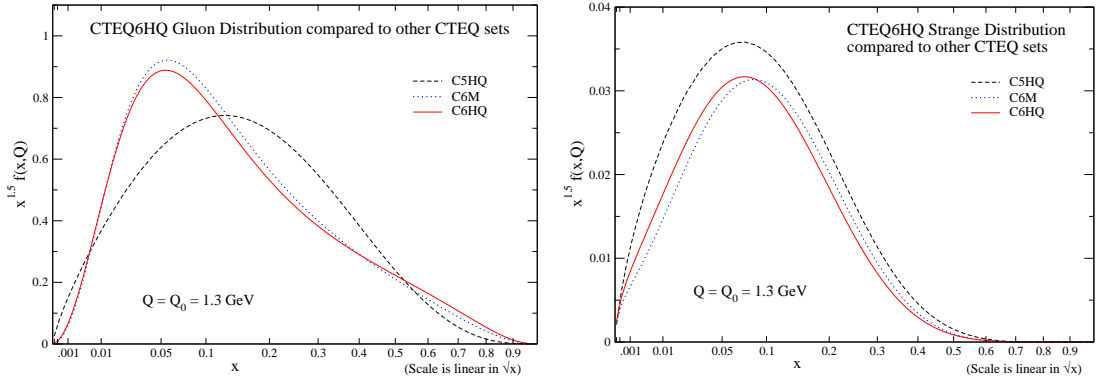

\includegraphics[%
  clip,
  width=0.48\columnwidth,
  keepaspectratio]{eps/GluQ0A.eps}~~~\includegraphics[%
  clip,
  width=0.48\columnwidth,
  keepaspectratio]{eps/StrQ0A.eps}

\caption{Comparison of the a) gluon and b) strange-quark distributions at
$Q_{0}=m_{c}=1.3\:$GeV for the CTEQ5HQ, CTEQ6M, and CTEQ6HQ sets.
The axes are scaled to highlight the valence components of these distributions.\label{figone}}
\end{figure}
\begin{figure}
\includegraphics[%
  clip,
  width=0.48\columnwidth]{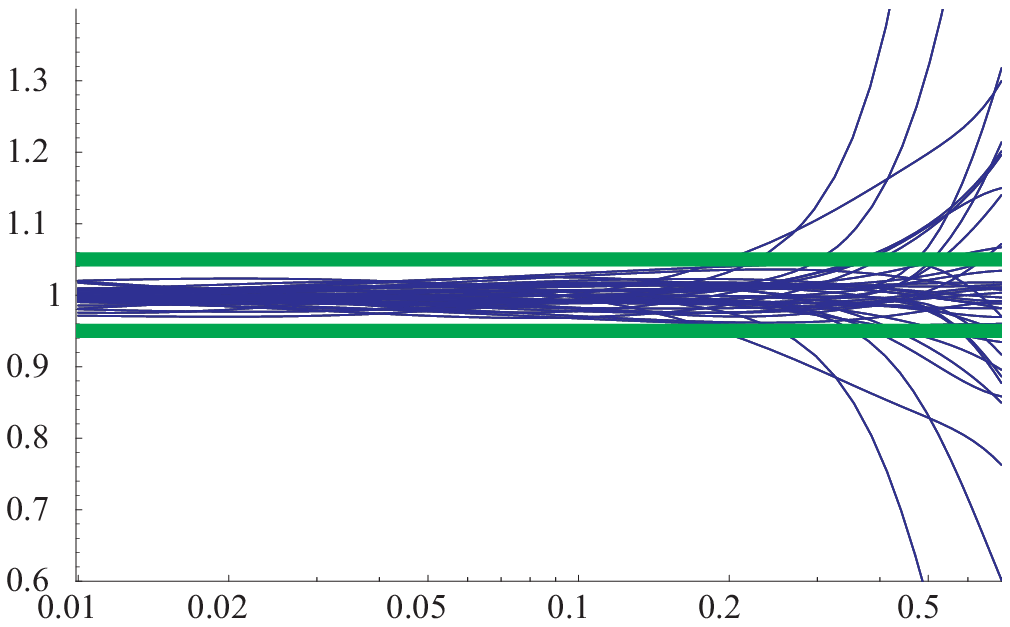}\includegraphics[%
  width=0.48\columnwidth,
  keepaspectratio]{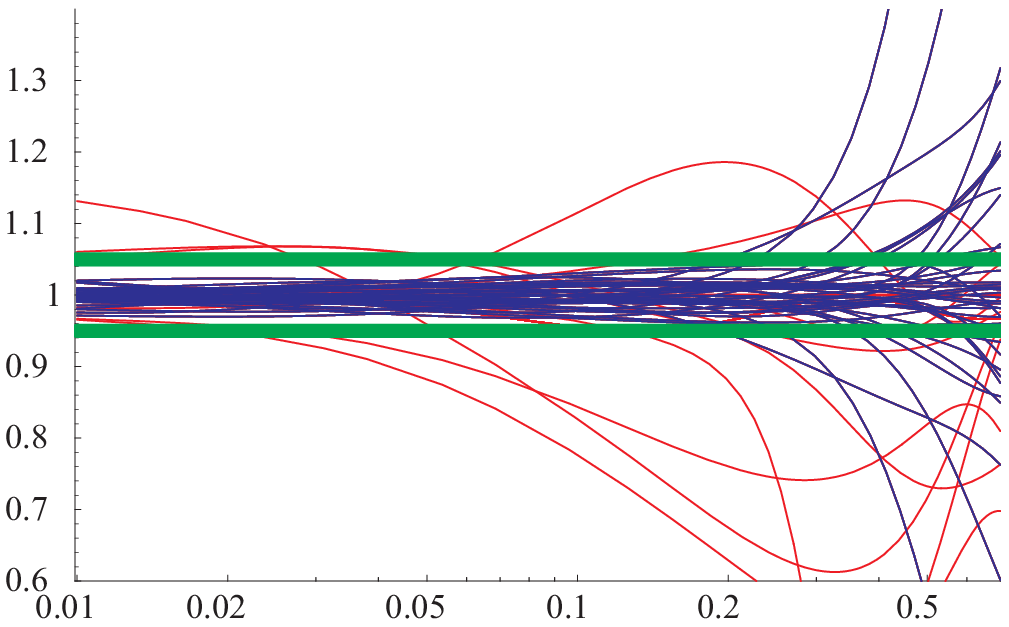}

\caption{a)~Ratio of the 40 CTEQ6M PDF sets compared to the central set for
the strange-quark vs~$x$. b)~Same as previous, with additional
sets included (with a more general $s(x)$ parameterization). In both
figures, guide-lines are drawn at $\pm5\%$.\label{figtwo}}
\end{figure}

\subsubsection{\textsc{Comparison with related PDFS: }}

The C6HQ and C6M fits provide comparable descriptions of the global
QCD data in two \textit{different} schemes. Some of the differences
in the PDFs arise purely from the choice of scheme. We are particularly
interested in the differences for the gluon distribution, which will
strongly influence the closely correlated charm distribution (since
it is generated via the $g\to c\bar{c}$ process). It is also interesting
to compare the differences between the earlier CTEQ5HQ (C5HQ) distributions
with the new C6HQ distributions; differences between these PDFs are
attributable both to new data, and to minor differences in the way
the theoretical inputs are implemented.

Fig.~\ref{figone}a) shows the comparison of the gluon distribution
at $Q_{0}$. Here the difference between C5HQ and the CTEQ6 generation
of gluon distributions is pronounced. The change in this least-well-determined
parton distribution is due to the recent precision DIS data (most
influential in the small $x$ region) in conjunction with the greatly
improved inclusive jet data from the Tevatron (critical for the medium
to large $x$ regions). The differences between C6HQ and C6M gluons
at large $x$ are due to a combination of scheme-dependence, and the
inherent uncertainty range of the current analysis.

Fig.~\ref{figone}b) shows the comparison of the strange distributions
at the same $Q_{0}$. The noticeable difference between the C5HQ curve
and the others, in this case, is largely the result of different theoretical
inputs: namely, the $\kappa$ parameter which determines the ratio
of strange to non-strange sea quarks at the initial scale $Q_{0}$.
This $\kappa$ factor, known only approximately, was chosen to be
$1/2$ both in the CTEQ5 and CTEQ6 analyses, but for slightly different
values of $Q_{0}$: 1.0~GeV for C5HQ, and 1.3~GeV for the CTEQ6
sets. We now look at the uncertainty of the strange quark in further
detail; this has received increased attention recently since this
is an important ingredient in the NuTeV determination of $\sin\theta_{W}$.

\subsubsection{\textsc{Strange-quark PDF and uncertainties:} }

Using the set of 40 C6M PDF sets, we can produce a band of of distributions
which, in principle, should characterize the uncertainty of the s-quark,
cf., Fig.~\ref{figtwo}a). Based on this plot, one would be tempted
to conclude that the uncertainty on the strange quark is better than
5\% over much of the $x$ range. Because the parameterization of $s(x)$
in the global fit is constrained to be $\kappa[\bar{u}(x)+\bar{d}(x)]$
at $Q_{0}$, the figure actually reflects the uncertainty not on $s(x)$
but instead on $[\bar{u}(x)+\bar{d}(x)]$. This constraint is imposed
because none of the data in the global analysis directly measures
the $s(x)$ distribution; hence, Fig.~\ref{figtwo}a) is by no means
a representation of the true uncertainty. 

Recent measurements of the charged-current charm production ($\nu s\rightarrow c\mu\rightarrow\mu^{\pm}\mu^{\mp}X$)
from CCFR and NuTeV \cite{Bazarko:1994tt,Goncharov:2001qe} have the
potential to determine $\{ s(x),{\bar{s}}(x)\}$ with much more precision.
In Fig.~\ref{figtwo}b) we again show the 40 C6M PDF sets combined
with a number of additional fits that relax the parameterization of
$s(x)$. This is not an exhaustive collection designed to span the
full parameter space, but rather an illustration that the implied
uncertainty of Fig.~\ref{figtwo}a) is much too conservative; the
dimuon data will go a long way toward improving our knowledge of the
strange PDF. An accurate determination of the strangeness of the proton,
as well as the strangeness asymmetry $[s(x)-{\bar{s}}(x)]$ will have
important implications for a number of measurements, including the
NuTeV $\sin\theta_{W}$ measurement.

\subsubsection{\textsc{Concluding Remarks:} }

The C6HQ PDFs presented here complement the CTEQ6 sets by providing
distributions which can be used in the generalized \msbar\  scheme
with massive partons. This analysis includes the complete set of NLO
processes including the real and virtual quark-initiated terms. 

While the zero-mass parton scheme is sufficient for many purposes,
the fully massive scheme can be important when physical quantities
are sufficiently sensitive to heavy quark contributions. This is evident
when comparing the C6HQ and C6M fits to the mis-matched sets where
the precise DIS data from HERA highlights the discrepancies.

The C6HQ fits also provide the basis for a series of further studies
involving more quantitative analysis of strange, charm, and bottom
quark distributions inside the nucleon. For example, the C6HQ PDFs
are necessary for a consistent analysis of resumed differential distributions
for heavy quark production.\cite{Nadolsky:2002jr} Using the full
range of data from both the charged and neutral current processes,
these distributions can reduce the uncertainties in the calculations;
hence, they have significant implications for charm and bottom production,
and can help resolve questions about intrinsic heavy quark constituents
inside the proton, the $\Delta xF_{3}$ structure function, and the
extraction of $\sin\theta_{W}$.

\subsubsection{\textsc{Acknowledgments}: }

F.I.O. acknowledges the hospitality of Fermilab and  BNL, where a
portion of this work was performed. This work was supported by RIKEN,
BNL, the U.S.~DoE under grant DE-AC02-98CH10886 \& DE-FG03-95ER40908,
and the Lightner-Sams  Foundation. 


\bibliographystyle{aipproc}
\bibliography{hq}

\end{document}